\documentclass[preprint,12pt]{elsarticle}

\usepackage{graphicx}
\usepackage{epsfig}

\usepackage{amssymb}

\journal{Nuclear Physics B}

\begin{document}

\begin{frontmatter}



\title{On fictitious flux tubes associated to
composite fermions in 2D Hall systems}


\author{J. Jacak, L. Jacak}

\address{Institute of Physics, Wroclaw University
of Technology,
Wyb. Wyspia\'nskiego 27, 50-370 Wroc{\l}aw,
Poland}

\begin{abstract}

Cyclotron braid subgroups are defined in order to
identify the topological origin
of Laughlin correlations in 2D Hall systems.
Flux-tubes and vortices for
composite fermion constructions are explained in
terms of cyclotron braids. A possible link of braid 
picture with quantum dynamics is conjectured in order to support 
the phenomenological model of composite fermions with auxiliary flux-tubes, for Landau level  fillings out of $\frac{1}{p}$, $p$ odd. A version of hierarchy for
fractional quantum  Hall effect is  proposed by mapping onto integer effect within the cyclotron 
braid approach. The  even
denominator
fractional lowest Landau level fillings, including Hall metal at $\nu=\frac{1}{2}$,  are also  
discussed in cyclotron braid terms.
\end{abstract}

\begin{keyword}

 Laughlin correlations, composite fermions, braid groups, Hall systems
\end{keyword}

\end{frontmatter}


\section{Introduction}
\label{}
In order to describe correlations in 2D charged
multi-particle systems
in the presence of strong perpendicular magnetic
field,
  the famous Laughlin
wave-function (LF) was introduced
\cite{laughlin2}.
The  representation  of the Coulomb interaction in terms of
the so-called
Haldane pseudopotential
 allowed for an observation \cite{haldane,prange,laughlin1} that
the LF exactly describes the ground state for $N$
charged 2D particles at the fractional Landau
level (LL) filling $1/q$, $q$--odd integer,
if one neglects the long-distance part of the
Coulomb
      interaction expressed by a
projection on the relative angular momenta of
particle pairs for values greater than $q-2$.
The Laughlin correlations were next effectively
modeled by
composite fermions (CFs) \cite{jain} in terms of
auxiliary flux-tubes
attached to particles.  By
virtue of the Aharonow-Bohm effect, the flux-tubes attached to particles produce
the required by LF statistical phase shift when
particle interchange.
The great advantage of the CF construction was
recognized in
possibility of interpretation of a fractional
quantum Hall effect (FQHE) in an external magnetic
field
as an integer quantum Hall effect (IQHE) in
resultant field screened by averaged field of the
fictitious flux-tubes \cite{jain}.
This allowed for recovery of the main line of
FQHE filling factor hierarchy,  $\nu=\frac{n}{(p-1)n\pm
1}$, ($p$--odd integer, $n$--integer) \cite{jain}, corresponding
to complete filling of $n$ LLs in the screened
field assuming that resultant field can be
oriented along or oppositely to the external
field (thus $\pm$ in the obtained hierarchy).
         Despite of
a wide practical usage of CFs in description of 2D
Hall systems, the origin and nature of
attached to particle flux-tubes are unclear,
similarly as  unclear is also the heuristic assumption that the
resultant
field screened by the mean field of local fluxes
can be oriented oppositely to the external
field (allowing, in that manner, for  the sign  minus in the hierarchy formula obtained by 
mapping of FQHE onto IQHE).

The competitive construction of CFs was also
formulated utilizing so-called
vortices \cite{vor3,vor1}, collective fluid-like
objects (in analogy of vortices in superfluid
systems)
that are assumed to be pinned to bare fermions and
reproducing Laughlin correlations \cite{vor3}.
Both types of composite particles, with vortices or with flux
tubes, are phenomenological in
nature, thus the question arises as to what
is a more fundamental reason of Laughlin correlations  in 2D charged
systems.

\begin{figure}[h]
\centering
{\includegraphics{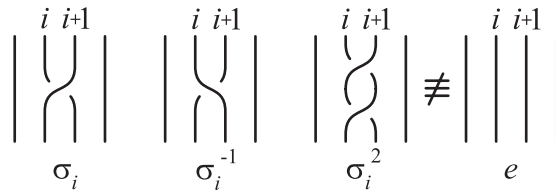}}
\caption{\label{fig:1} The geometrical presentation of the generator $\sigma_i$ of the full braid group for $R^2$ and its inverse $\sigma_i^{-1}$ (left); in 2D $\sigma_i^2\neq e$ (right)}
\end{figure}

It is well known \cite{wilczek,wu,sud},
that the source of exotic Laughlin correlations  is
of a 2D peculiar topology-type. This special
topology of planar systems is linked with
exceptionally rich structure of braid groups for 2D
manifolds (like $R^2$, or compact manifolds like sphere or torus) in
comparison to braid groups
for higher dimensional spaces ($R^d$, $d>2$)
\cite{birman}. The full braid group is defined as 
                       $\pi_1$ homotopy  group 
of the $N$--indistinguishable--particle configuration space, i.e.,  the group of  multi-particle 
trajectory classes, disjoint and  topologically nonequivalent (trajectories from various classes 
cannot be continuously transformed one onto another one).
The full braid groups is
infinite for 2D case while is finite (and equal to
the ordinary permutation
group $S_N$) in higher dimensions of the manifold
on which particles are located \cite{birman,jac}. This property makes two dimensional systems 
 exceptional in geometry--topology sense.
For matching  the topological properties with quantum
system properties, the
quantization according to the  Feynman path integral method
is particularly useful
\cite{wilczek,wu,lwitt}.
Due to a fundamental ideas of path integral
quantization in the case of not simply connected
configuration spaces (indicated by nontrivial $\pi_1$ group),
like  for the  multi-particle systems,  additional phase
factors---weights of nonequivalent
(nonhomotopic) trajectory classes and summation
over these classes must be included (a measure in the trajectory space is
distributed over  separated disjoint  homotopy classes of $\pi_1$). As it was proved in
\cite{lwitt}, these weight factors form a
one-dimensional unitary representation (1DUR) of
the
full braid group. Different 1DURs of the full
braid group give rise to distinct types of quantum
particles corresponding to
the same classical ones. In this manner one can
get fermions and bosons corresponding to only
possible 1DURs of $S_N$, 
$\sigma _i \rightarrow e^{i\pi}$ and
$\sigma_i\rightarrow e^{i0}$, respectively,
 (the permutation group $S_N$ is the full braid group in 3D and in higher dimensions,
$\sigma_i$, $i=1,...,N$ denote  generators of $S_N$).
For more rich  braid group in 2D one encounters,
however,
the infinite number of possible so-called anyons (including
bosons an fermions)
related to 1DURs, $\sigma_i\rightarrow
e^{i\Theta}$, $\Theta \in [0,2\pi)$ ($\sigma_i$
are here generators
of the full braid group in 2D, cf. Fig. \ref{fig:1})
\cite{wilczek,wu,sud,birman}. 

CFs associated with
Laughlin correlations require, however, the statistical  phase shift
$p \pi$, with $p=3,5,7...$ for its various types,
and  the periodicity od 1DURs,
$e^{ip\pi}=e^\pi=-1$,
does not allow to distinguish CFs from ordinary
fermions. It caused some misinterpretation---CF
fermions were treated \cite{jain,hon} as ordinary
fermions dressed somehow with flux-tubes in
analogy to solid-state quasiparticles, which is,
however, an  incorrect picture.



In the present paper we revisit the topological
approach to Hall systems and recover Laughlin correlations  by
employing properties of the underlying cyclotron
braids \cite{jac1,jac2}, originally defined
and
without a phenomenological modeling of CFs. 
We will
demonstrate that particles with
statistical properties of CFs are not composites
of fermions with flux-tubes
or vortices, but are rightful  2D quantum
particles characterized by 1DURs of cyclotron
braid subgroups. We notice
 also that
the original CFs construction with flux-tubes employing 
a heuristic assumption that 
the mean field of local fluxes can be greater than
the external, to be justified in terms of cyclotron 
braid subgroups needs a special assumption on possible quantum dynamics.  One can formulate, however,  the
  recovery of FQHE hierarchy in terms of IQHE
in resultant field, avoiding the
previously made  assumption on possible $\pm$ sign of the effective field.
The explanation  of the mechanism for creation of the effective field
for fractional fillings of Landau level (LL) in
terms of cyclotron braid groups would be also helpful
for
identification of Chern-Simons field constructions
\cite{cs}, which were widely spread for modeling
of CFs and anyons within mathematical effective
approach to Hall systems in fractional regime.


\section{Too-short for interchanges cyclotron
trajectories in 2D Hall systems}

One-dimensional unitary representations (1DURs) of
full braid group \cite{wu,birman,imbo}, i.e.,
of $\pi_1$ homotopy group of the 
configuration space for indistinguishable
$N$ particles  \cite{birman},
 define weights for the
path integral summation over trajectories
\cite{,wilczek,wu,lwitt}.
If trajectories fall into separated homotopy
classes that are distinguished by non-equivalent
closed loops (from $\pi_1$) attached to an open trajectory
$\lambda_{a,b}$ (linking in the configuration
space points, $a$ and $b$),
then an additional summation over these classes
with an appropriate
unitary factor (the weight of the particular
trajectory class) should be included
\cite{wilczek,wu} in the path
        integral (for  transition from the point $a$ at the time moment $t=t_1$ to the point $b$
at $t=t_2$):
         \begin{equation}
I_{a,t_1\rightarrow b,
t_2}=\sum\limits_{l\in\pi_1} e^{i\alpha_l}\int
d\lambda_l e^{iS[\lambda^{l}_{(a,b)}]},
         \end{equation}
          where
       $\pi_1$ stands
         for the full braid group and index $l$ enumerates $\pi_1$ group elements, $\lambda^l$ indicates
an open trajectory $\lambda$ with added  $l$th loop from $\pi_1$ (the full  braid group here).
The factors $ e^{i\alpha_l}$ form a 1DUR of the
full braid group and distinct representations
correspond to distinct
types of quantum particles \cite{wu,lwitt}.
The closed loops from the full braid group
describe exchanges of identical particles, thus,
the full braid group 1DURs
indicate the statistics of  particles
\cite{wilczek,wu,sud}.

Nevertheless, it is impossible to associate in
this manner CFs with the 1DURs of the full braid
group, because 1DURs are periodic with a period of
$2\pi$, but CFs
require the statistical phase shift of $p\pi$,
$p=3,5...$. In order to solve this problem,
 we propose \cite{jac1} to associate CFs with
 appropriately constructed braid
subgroups instead of the full braid group and in
this way to distinguish CFs from ordinary
fermions.

The full braid group contains all accessible
closed multi-particle classical trajectories,
i.e., braids
(with  initial and final orderings of particles
that may differ by permutation, which is admitted for indistinguishable particles).
One can, however, notice that inclusion of a
magnetic field substantially changes
trajectories---a classical cyclotron motion
 confines a
  variety of accessible braids.
When the separation of particles is greater than
twice the cyclotron radius, which situation occurs at
fractional lowest LL fillings,
the exchanges of particles along single-loop
cyclotron trajectories are {\it precluded},
because the cyclotron orbits are {\it too short}
   for particle  interchanges.
Particles must, however, interchange in the braid
picture for defining the statistics and in order
to allow exchanges again, the cyclotron
radius must somehow be {\it enhanced}. The natural
way is to exclude inaccessible braids from the
braid group. We will show   that remaining braids would be
sufficient for particle  exchanges realization.

One can argue that cyclotron radius enhancement
could be achieved by either lowering the effective
magnetic field or
lowering the effective particle charge. These two
possibilities lead to the two phenomenological
concepts of
CFs---with the lowered field in Jain's
construction \cite{jain} and with the screened
charge in Read's construction of vortices
\cite{vor3}.
Both these constructions seem to not matter with
braid groups, but actually
both of these effective phenomenological tricks
correspond to the same, more basic and natural
concept, of restricting the braid family by
excluding inaccessible trajectories
\cite{jac1,jac2}. We will demonstrate below that at
sufficiently high
magnetic fields in 2D charged $N$--particle
systems, the {\it multi-looped braids} allow for the
effective
enlargement of cyclotron orbits, thus restoring
particle exchanges in a natural way \cite{jac2}.
These multi-looped braids form a subgroup of the
full braid group
and, in the presence
of strong magnetic field, the summation in the
Feynman propagator will be thus confined to the
elements of this subgroup (its semigroup,
for fixed magnetic field orientation, however,
with the same 1DURs as of the subgroup).

\section{Cyclotron braid subgroups---restitution of particle interchanges}

\begin{figure}[h]
\centering
{\includegraphics{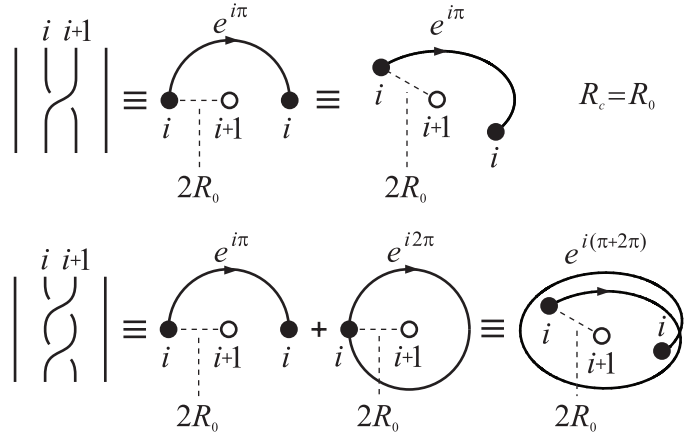}}
\caption{\label{fig3} The generator $\sigma_i$ of the full braid group and the corresponding relative trajectory of the $i$th and $(i+1)$th particles exchange (upper); the generator of the cyclotron braid subgroup, $b_{i}^{(p)}=\sigma_i^p$ (in the figure, $p=3$), corresponds to additional $\frac{p-1}{2}$ loops when the $i$th particle interchanges with the $(i+1)$th one  (lower) ($2R_0$ is the inter-particle separation, $R_c$ is the cyclotron radius, 3D added for better visualization)}
\end{figure}

Mentioned above multi-looped  braids form the
   {\it cyclotron braid subgroups} and are
 generated by the following generators:
\begin{equation}
\label{gen}
b_i^{(p)}=\sigma_i^p,
\;\;(p=3,5...),\;i=1,...,N-1,
\end{equation}
where each $p$ corresponds to a different type of
the cyclotron subgroup
and $\sigma_i$ are the generators of the full
braid group.
The group element $b_i^{(p)}$ represents the
interchanges of the $i$th and $(i+1)$th particles
with $\frac{p-1}{2}$ loops,
which is clear by virtue of the definition of the
single interchange $\sigma_i$ (cf. Fig.
\ref{fig3}, e.g., for $p=3$ one deals with elementary particle exchange braid with one additional loop).
 It is clear that $b_i^{(p)}$ generate a subgroup of the full braid
group as they are expressed by the full braid group  generators
$\sigma_i$.


The 1DURs of the full group confined to the
cyclotron subgroup (they
do not depend on $i$ as 1DURs of the full braid
group do not depend on $i$ by virtue of the $\sigma_i$
generators property,
$\sigma_i\sigma_{i+1}\sigma_i=\sigma_{i+1}\sigma_i
\sigma_{i+1},\;\; 1\leq i\leq N-1$,
\cite{birman,jac}) are 1DURs of the cyclotron
subgroup:
\begin{equation}
\label{repr}
b_i^{(p)}\rightarrow e^{ip\alpha},\;i=1,...,N-1,
\end{equation}
where $p$ is  an odd integer and
$\alpha \in (-\pi,\pi]$. We argue, that these 1DURs, enumerated
by the {\it pairs} ($p$, $\alpha$),
describe composite anyons (CFs, for $\alpha
=\pi$). Thus in order to distinguish
various types of composite particles one has to
consider ($p$, $\alpha$) 1DURs of cyclotron braid
subgroups.

In agreement with the general rules of
quantization \cite{sud,imbo}, the $N$-particle
wave function must transform according to the 1DUR
of an appropriate element of the braid group, when
the particles traverse, in classical terms, a
closed loop in the configuration space
corresponding to
this particular braid element. In this way the
wave function acquires
 an appropriate phase shift due to
particle interchanges (i.e., due to exchanges of
its variables according to  the prescription given by
braids in 2D configuration space).
Using 1DURs as given by (\ref{repr}), the
Aharonov-Bohm phase of Jain's fictitious
fluxes is replaced by contribution of additional
loops (each loop adds $2\pi$ to the total phase
shift, if one considers
1DUR with $\alpha=\pi$ related to CFs, cf. Fig.
\ref{fig3} (right)).
Let us emphasize that the real particles do not
traverse the braid trajectories, as quantum
particles do not have any trajectories,
but exchanges of coordinates of the $N$-particle wave
function can be represented by braid group
elements; in 2D a coordinate exchange do not
resolve itself to permutation only, as it was
in 3D, but must be performed according to an appropriate element 
of the braid group, being in 2D not the same as the permutation group  \cite{wu,sud,imbo}.
Hence, for the braid cyclotron subgroup generated
by $b_i^{(p)}$, $i=1,...,N-1$, we obtain
the statistical phase shifts $p\pi$ for CFs
(i.e., for $\alpha=\pi$ in Eq. (\ref{repr})), as
required by Laughlin correlations, without the need to model them with flux
tubes or vortices.

\begin{figure}[h]
\centering
\scalebox{0.7}{\includegraphics{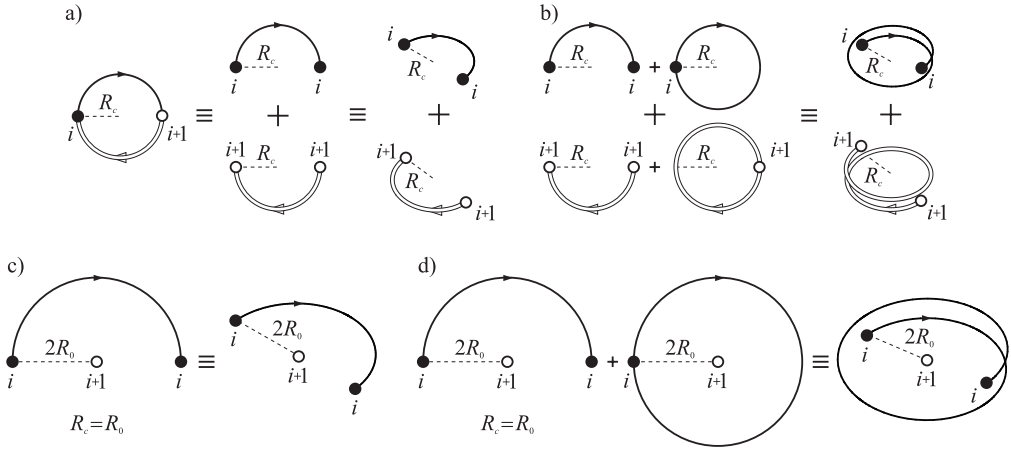}}
\caption{\label{fig4} Half of the individual particle cyclotron trajectories of the $i$th and $(i+1)$th particles (top) and the corresponding relative trajectories (bottom) for interchanges of the $i$th and $(i+1)$th 2D-particles under a strong magnetic field, for $\nu =1$ (left) and for $\nu =\frac{1}{3}$ (right), respectively ($R_c$---cyclotron radius, $2R_0$---particle separation, 
3D added for better visualization)}
\end{figure}

\begin{figure}[h]
\centering
\scalebox{0.7}{\includegraphics{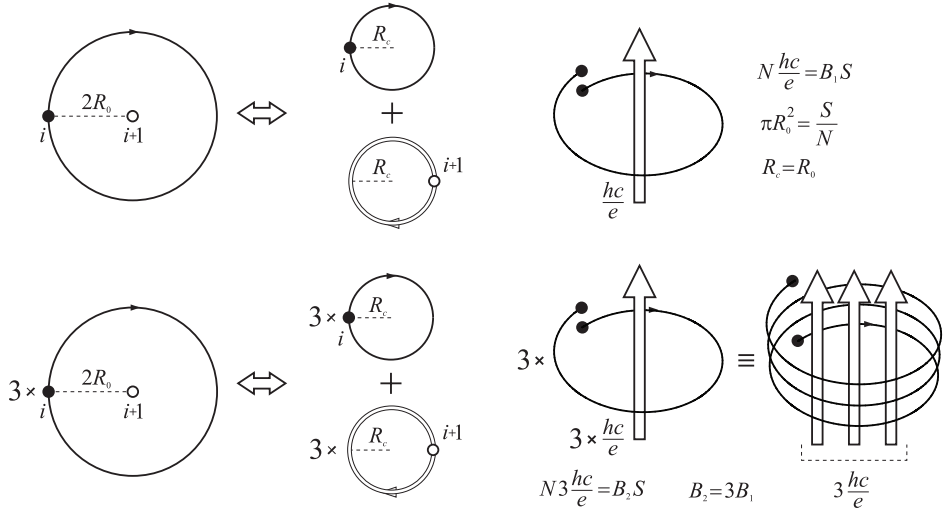}}
\caption{\label{fig5} Cyclotron trajectories of individual particles must be closed, therefore they correspond to 
{\it double} exchange braids,  for both,  simple exchanges (upper) and  exchanges with additional loops (lower), in the right part, quantization
of flux per particle, for $\nu=1$ and $\nu=\frac{1}{3}$, is indicated}
\end{figure}

Each additional loop of a relative trajectory for
the particle pair interchange
(as defined by the generators $b_i^{(p)}$)
reproduces
an additional loop in the individual cyclotron
trajectories for both interchanging
particles---cf. Fig. \ref{fig4}.
The cyclotron trajectories are repeated in the
relative trajectory (c,d) with twice the radius of
the
individual particle trajectories (a,b). In quantum
language, with regards to classical multi-looped
cyclotron
trajectories, one can conclude only on the number,
$\frac{BS}{N}/\frac{hc}{e}$, of flux quanta per
single
particle in the system, which for the filling
$\frac{1}{p}$ is $p$ (for odd integer $p$), i.e.,
the same as the number of individual particle
cyclotron loops (which equals to $p=2n+1$, where
$n=1,2...$ indicates the number of additional
braid-loops for particle interchange trajectories). From this observation it follows a
simple rule: for $\nu =\frac{1}{p}$ ($p$ odd), each
additional loop of a
cyclotron braid corresponding to particle
interchange, results in {\it two} additional flux
quanta
piercing the individual particle cyclotron
trajectories.
This rule follows immediately from the definition
of the cyclotron trajectory, which
must be a {\it closed} individual particle
trajectory related to
a {\it double} interchange of the particle pair (cf. Fig. \ref{fig5}).
In this way, the cyclotron trajectories
of both interchanging particles are closed, just
like the closed relative trajectory for the {\it
double} interchange
(the braid trajectory is open as the trajectory of
 particle interchange only, and therefore  the {\it
double} interchange is needed to close this trajectory).
If the interchange is simple, i.e., without any
additional loops, the corresponding individual
particle cyclotron trajectories are also simple,
i.e., single-looped. Nevertheless, when the
interchange of particles is multi-looped, as
associated with the $p$-type cyclotron subgroup
($p>1$), the
double interchange relative trajectory has $2
\frac{p-1}{2}+1=p$ closed loops, and the
individual cyclotron
trajectories are also multi-looped, with $p$ loops
\cite{jac2,jac-epl}.


All these properties of multi-looped planar
trajectories   at strong magnetic field are linked with
the fact that in 2D additional loops cannot enhance
the total surface of the system.
In this regard,  it is important to emphasize the basic
difference between the turns of a 3D winding
(e.g., of a wire)
and of multi-looped  2D cyclotron trajectories. 2D
multi-looped trajectories do not enhance the surface
of the system and therefore do not enhance
total magnetic field flux $BS$ piercing the
system, in opposition to  3D case. In 3D case,
each turn
of a winding adds a new portion of  flux, just
as a new turn adds a new surface, which is,
however, impossible in 2D.
Thus in 2D all loops must share the same total
flux, which results in {\it diminishing} flux-portion
per a single loop and, effectively,
in longer cyclotron radius (allowing again
particle interchanges).

The additional loops in 2D take away the
flux-portions (equal to $p-1$  flux quanta just at
$\nu=\frac{1}{p}$, $p$ odd) simultaneously
diminishing the effective field; this gives an
explanation
for Jain's auxiliary fluxes screening the external
field $B$.
Thus, it is clear that CFs are actually not
compositions of particles with flux-tubes, but are
rightful particles in 2D
corresponding to 1DURs of the cyclotron subgroups
instead of the full braid group, which is
unavoidably forced by too short ordinary
single-looped cyclotron trajectories.
The original  name 'composite fermions' 
can be, however, still used. Moreover, one can use
a similar name, 'composite anyons', for particles
associated with fractional 1DURs (i.e., with
fractional $\alpha$) of the cyclotron
subgroup instead of the full braid group, the latter  linked
rather with ordinary anyons (without magnetic
field).

\section{The mapping of  FQHE onto IQHE}

Let us emphasize that the agreement between loop
number and flux quanta number per particle
(allowing, in fact, for a Jain' model of CFs)
is restricted only to fillings $\frac{1}{p}$, $p$
odd. Out of these fillings, the number of flux
quanta fer particle cannot be equal to number of
loops
as it is not an integer (while the number of loops is
always integer). For fields out of
$\nu=\frac{1}{p}$,
on a single loop it falls not integer number of flux
quanta (it may happen, because cyclotron loops are
classical
braid-type objects, not quasiclassical trajectories
with flux quantization requirements).
 In other words, all loops together
must take away the total flux of the external field.
Oppositely it is assumed in
CF construction with concept of
rigid flux quanta attached to particles even out
of  $\nu=\frac{1}{p}$   filling fraction. In particular, it is assumed by Jain
that the resultant effective field of the fictitious flux quanta and of the
external magnetic field would be even negative (oriented oppositely to the external filed) as e.g., for $\nu >\frac{1}{2}$ (for $p=3$) 
 \cite{jain,hon}. 
The concept of Jain's resultant field leads to   the FQHE hierarchy obtained via mapping of FQHE onto 
IQHE, $\nu=\frac{n}{(p-1)n\pm 1}$ \cite{jain}.

 One can, however, consider the
mapping of FQHE onto IQHE within cyclotron braid approach, assuming uniform distribution of the external field flux over all trajectory loops.
From  point of view of
multi-looped
cyclotron braids, in the case of $\nu\neq
\frac{1}{p}$ ($p$ odd), on each loop it falls a
fraction of a flux quantum and if it coincides
with the same fraction as per single particle for
 completely filled several Landau levels (with
single cyclotron loops) the mapping of
IQHE onto FQHE holds, resulting in filling
hierarchy. One can compare the flux-fractions per single loop, for fractional and integer 
LLs fillings:
\begin{equation}
\begin{array}{llll}
FQHE: & \nu=\frac{N}{N_0},& N_0=\frac{BS}{hc/e},&
\Phi_F=\frac{BS}{Np}=\frac{hc}{e\nu p},\\
IQHE (n-th\;\; LL):& n=\frac{N_1}{N_0},&
N_0=\frac{B_1S}{hc/e},& \Phi_I=\frac{B_1
S}{N_1}=\frac{hc}{en},\\
\end{array}
\end{equation}
and, in the case when the flux per single loop in
FQHE, $\Phi_F
=\frac{BS}{Np}=\frac{hc}{e\nu p}$ is equal to the
flux per single particle  in IQHE (thus, per single loop, as for IQHE cyclotron trajectories are single-looped),
$\Phi_I=\frac{B_1 S}{N_1}
=\frac{hc}{en}$, the mapping of FQHE onto IQHE
holds, and it happens when
$\nu=\frac{n}{p}$,  where $n=1,2,3,4...$,
$p=1,3,5...$. This  reproduces FQHE
hierarchy (due to the
mapping onto IQHE) avoiding
problems with  sign minus in the former
formula $\nu=\frac{n}{(p-1)n\pm 1}$ \cite{jain}.
 For  filling rates out of $\frac{1}{p}$, $p$--odd, one deals
with still integer number of additional loops
per particle but not with  integer number of flux quanta.

Any flux-tubes  attached to composite fermions do not exist, they are only a
convenient model for
additional loops (allowing for  interchanges when
single-looped cyclotron trajectories are too short) and
exceptionally for $\nu=\frac{1}{p}$ ($p$--odd)
they would be imagined as of $p-1$ flux quanta attached to particles and oppositely oriented 
 to external field, 
but out of these fillings, not.
In order to rescue, however, the assumption of Jain, that even out of $\nu=\frac{1}{p}$, to particles are associated complete flux quanta one can consider the situation when particular loops of multi-loop structure embrace the 
integer number of flux quanta and only the last one takes a fractional rest. 
This needs, however, the conjecture that the braid multi-looped structure is 
repeated by semiclassical quantum dynamics of wave-packets. Formation of such wave-packets traversing closed cyclotron loops (in order to define flux of field trough their  orbits) is highly sophisticated, as single particle quantum dynamics might not manifest cylindric symmetry (depending of a gauge choice) \cite{landau1972}, and operators corresponding to classical position of cyclotron orbit center do not commute \cite{eliu}. Nevertheless,  for quantum evolution in magnetic field one can write out a cyclic in time position and momentum operators in Heisenberg picture and thus, periodic (with cyclotron period) evolution of any wave-packet \cite{eliu}. 

 In the case when distributing a total flux quantum over all loops and the rest per the last loop  is equal to $n$-th part of the flux quantum (with sign minus or plus), one can expect organization of quantum dynamics in  the form of orbital movement of wave-packets representing single particles and thus forced rigid quantization of  the external magnetic field flux passing trough the surfaces of these orbits. Interesting is the last loop taking only $\pm \frac{hc}{ne}$ flux, with its sign depending of the orientation of the resulting Jain's field with respect to the external field. In order to fulfill, however, qusiclassical rule of flux quantization one can imagine organization of the collective $n$ particle closed trajectory embracing flux quantum in analogy to quantum dynamics of completely filled $n$ Landau levels. Here one can invoke a pictorial interpretation of the cyclotron trajectory fitted in length to $n$ de Broglie waves.

The negative  sign of the rest flux passing through the last loop  means here the opposite direction (with respect to remaining loops, contrary to the direction induced by the external field) movement, resulting in the something like eight-shape multi-looped cyclotronic structure of wave-packet trajectory. Such strange picture of possible arrangement of quasiclassical movement for this particular LL fillings (given by Jain's FQHE hierarchy $\nu=\frac{n}{n(p-1)-1}$) satisfies, however, requirements of flux quantization and explains the heuristic assumption of Jain's composite fermion construction. 

With regard to the above formulated conjecture interesting would be a measurement of cyclotron focusing of tiny beam of 2D carriers passing trough a nanometer-scale slot. This beam  bent by magnetic field   to the left or to the right with respect to the source slot, for two directions of cyclotron movement, can be resonantly observed in nearby target slots in the case of commensuration  of the cyclotron radius and the separation of slots  \cite{ogn,hon}. An asymmetry should be observed when passing  $\nu=\frac{1}{2}$ LL filling by lowering or rising  the magnitude of the external field. 
This would be helpful in experimental verification of one of two above presented possibilities with either uniform distribution of the external field flux over all loops or nonuniform distribution forced by quasiclassical rule of flux quantization.

The other  problem raised by cyclotron
braid approach consists in the  fact that CF are not
ordinary fermions dressed with  interaction,
but are separated 2D quantum particles, and they cannot
be mixed with ordinary fermions (similarly as
bosons cannot be  mixed with fermions), especially
within numerical variational interaction
minimizations or diagonalizations. Even though both fermions and CFs
correspond to antisymmetric wave functions, not
all antisymmetric wave functions describe CFs and
the domain for minimization can comprise only
these antisymmetric functions which
transform according to appropriate 1DUR of
the cyclotron subgroup (it is a subspace of the
Hilbert space of antisymmetric functions). The
minimizations done on the whole domain of
antisymmetric functions would lead thus to
improper results
and should be repeated on the confined domain in
the Hilbert space.

\section{The influence   of the
Coulomb interaction}

The Coulomb interaction play a central role  for Laughlin correlations 
\cite{haldane,prange,laughlin1},  but in 2D systems upon the quantized  magnetic field, 
the interaction of charges  cannot be
accounted for
in a manner of standard dressing  of particles with
interaction as it was  typical for quasiparticles in
solids, because in 2D Hall regime 
this interaction does not have a continuous
spectrum with respect to particle separation
expressed in relative angular momentum
terms \cite{haldane,prange}. The interaction can
be operationally included within the Chern-Simons
(Ch-S) field theory
\cite{cs,lopez}, formulating an effective description of the
local gauge field attached to particles, which, in
the area of Hall systems,
suits to particles with vortices, such as anyons
and CFs \cite{hon}.
It has been demonstrated \cite{prange,mo} that the
short-range part of the Coulomb interaction
stabilizes CFs
against the action of the Ch-S field (its
antihermitian term \cite{mo,vor16}), which mixes
states with distinct angular momenta
within LL \cite{mo}, in disagreement with the CF
model in the Ch-S field approach \cite{hon,mo}.
The Coulomb
interaction removes the degeneracy of these states
and results in energy gaps which stabilize the CF
picture, especially
effectively for the lowest LL. For higher LLs, the
CFs are not as useful due to possible mixing
between the LLs
      induced by the interaction \cite{id}.
The short-range part of the Coulomb interaction
also stabilizes the CFs in cyclotron braid terms
\cite{jac1}, similarly to how
it removes the instability caused by the Ch-S
field for angular momentum orbits in LL \cite{mo}.
Indeed, if
the short-range part of the Coulomb repulsion was
reduced, the separation of particles would not be
rigidly
kept (adjusted to a density only in average) and
then other cyclotron trajectories, in addition to
those
for a fixed particle separation (multi-loop at
$\nu=\frac{1}{p}$), would be admitted, which would
violate the cyclotron subgroup construction.

\section{Read's CFs and Hall metal state in cyclotron braid terms }

For Read's CFs \cite{vor3,vor1}, Laughlin correlations are modeled
by collective vortices that are attached to the
particles.
A vortex with its center at $z$ is defined as
\cite{vor3},
\begin{equation}
 \label{vor1}
 V(z)=\prod_{j=1}^{N} (z_j-z)^q,
\end{equation}
where $q$ is the vorticity. For odd $q$, it is
linked to the Jastrow factor
of the LF \cite{laughlin2}, $\prod_{i>j}^{N} (z_j-z_i)^q$, (resulting from Eq. (\ref{vor1}) by the replacement of $z$
with $z_i$ and the addition of $i$ ($i>j$) to the
product domain,
 i.e., by the  binding of vortices
to electrons). In particular, for $q=1$ one
arrives
at the Vandermonde determinant,   $\prod_{i>j}^{N} (z_j-z_i) $ (being the polynomial  part  of the Slater function 
of $N$ noninteracting 2D fermions at magnetic field corresponding to $\nu=1$, i.e., to the case of the complete filled lowest LL), associated with
the ordinary single-looped cyclotron motion of $N$
fermions on the plane at $\nu=1$.
 Because the vortices are  fragments 
of the LF,  they contain more
information than just the statistical winding
phase shift
(the latter expressed by the factor,
$\prod_{i,j}(z_i-z_j)^q/|z_i-z_j|^q$).
 1DURs
of the cyclotron braid subgroups define the
statistical phase winding,
but not the shape of the wave function. 
The wave function shape  is
determined via the energy competition between
various
wave functions with the same statistical symmetry. Thus,
vortices contain
information beyond just the statistical phase
shift, they also include the specific radial
dependence
of multi-fold zeros pinned to particles through
the Jastrow polynomial.
The vortex
is a collective fluid-like concept that does
not meet the single-particle picture.
The vorticity $q$ is selected, however, in
accordance with the {\it known in advance}
LF, thus, similarly as
CF flux tubes, it requires a motivation within the
cyclotron structure.

The properties of vortices can be listed as
follows \cite{vor3}:
\begin{itemize}
\item{ when traversing with an arbitrary particle
$z_j$ a closed loop around
the vortex center, then the gain in phase is equal
to $2\pi q$;}
\item{the vortex induces a depletion of the local
charge density, which results in
a locally positive charge (due to background
jellium) that screens the charge of the electron
associated to the vortex center; this positive
charge is $-q\nu e$
(for $\nu =1/q$ it gives $-e$, which would
completely screen the electron charge);}
\item{exchange of vortices results in a phase
shift of $q^2\nu \pi$, (due to the charge deficit
of the vortex),
 which for $\nu=\frac{1}{q}$ gives $ q\pi$;
the $q$-fold vortex, together with the bound
electron (which contributes a charge $e$ to the
complex
and produces a statistics phase shift of $\pi$),
form a complex that
behaves like a composite boson with zero effective
charge for odd $q$ and like a composite fermion
 for even $q$.}
\end{itemize} 

The bosons can condense and, in this manner, 
reproduce exactly the LF for odd $q$ \cite{vor16},
while, for even $q$, one deals with the Fermi sea
in a zero net field, as in
both cases the effective charge of the complexes
is zero; the latter case corresponds to the Hall metal
state \cite{halperin,vor7,vor11}.

The second property of listed above,  explains why the model with
vortices works. The reduced effective charge
of the electron--vortex complex, results in an
increase of the cyclotron radius,
which is necessary for particle exchanges at
fractional fillings.

     The specific character
of the concept of vortices is clearly visible, in
particular, for $\nu=1$. Vortices of the form
(\ref{vor1}) with vorticity $q=1$, attached to
electrons
in the system, result in the Vandermonde factor
(being the Jastrow factor with exponential $q=1$).
In this case,
the corresponding Laughlin state is thus given by the Slater
function of $N$ {\it noninteracting}
fermions, what, however, can also be
effectively described by
     the Bose Einstein condensate of
     bosons 
defined as fermions with vortices  (\ref{transf}) with
$q=1$ (all the action of the magnetic field
on ordinary fermions is replaced with this Bose
condensation). The Coulomb interaction do not
contribute in
this particular case, since at $\nu=1$ the Haldane
pseudopotential \cite{haldane,prange} (i.e., the
short range part of the
     Coulomb interaction, being essential in the
selection of the Laughlin state form) is zero (as
$q-2<0$, for $q=1$), and thus the Slater function
of {\it noninteracting}
particles is suitable as the eigen-state
of the interacting system at $\nu=1$.

 A phenomenological modifications of
vortices, like a shift of the centre of the vortex
from the position of an associated electron, may
result in effective attraction
 of vortex-composite
fermions, leading to their pairing at e.g., $\nu
=5/2$ \cite{vor1,vor17}. This corresponds,
in fact, to a modification of the Laughlin
function
and leads to a new wave function, in this case,
$N$ particle BCS-like
 function
in the form of Pfaffian, as was described in Refs
\cite{vor1,vor17}. 
Note that the wave function with the Pfaffian
factor is still of the same statistical
symmetry as that for the particular sort of
braid-composite fermions (defined by 1DUR of the
corresponding
 cyclotron subgroup).

All properties of vortices or flux-tubes  in CF constructions can be grasped 
together by a formal local gauge transformation
\cite{vor16} of the original fermion particles
(defined by the  fermion 
field operator
   $\Psi({\bf x})$) to composite particles represented by fields (annihilation and creation):
  $
   \label{transf}
\Phi({\bf x})=e^{-J({\bf x})}\Psi({\bf x}),\;\;\Theta({\bf
x})=\Psi^+({\bf x})e^{J({\bf x})}$, where:
$J ({\bf x}) =q\int d^2x'\rho({\bf
x}')log(z-z')-\frac{|z|^2}{4l^2}$, and $e^{-J}
   $ corresponds to a nonunitary, in general,
    transformation that describes the attachment
of Read's vortices (or Jain's flux-tubes) to the
bare fermions, $ \Psi({\bf x})$ and $ \Psi^+({\bf
x})$
(for the original fermion annihilation and creation fields,
respectively).
When restricting $J ({\bf x})$ to only its
imaginary part (i.e., to the imaginary part of
$log$),
one arrives at the hermitian Ch-S field
corresponding to the dressing of fermions with
local flux-tubes
    \cite{vor18}. The field operators
$ \Phi({\bf x})$ and $ \Theta({\bf x})$,
$\Phi^+({\bf x})
=\Theta({\bf x})e^{J({\bf x})+J^+({\bf x})}$,
though are not mutually conjugated (they are
perfectly conjugated for the hermitian Ch-S
field),
      describe composite bosons (for odd $q$)
      and composite fermions
(for even $q$) within the mean field approach
\cite{vor16} (remarkably, the real part of $J$
vanishes in the mean field,
      as the real part of $log$
is canceled by the Gaussian, while the hermitian
Ch-S field is canceled by the external magnetic
field).
From the relation
$e^{q\sum_{j}log(z-z_j)}=\prod_{j}^{N}(z-z_j)^q$
(for the density operator $\rho({\bf
x})=\Psi^+({\bf x})\Psi({\bf x})\Longrightarrow
\sum_{j=1}^{N}\delta(z-z_j)$),
      which coincides with the definition
of Read's vortex, one can expect that the above
local gauge transformation reproduces all
properties of vortices.
This gauge transformation allows for the
interpretation of the Laughing state as
a Bose-Einstein condensate of composite bosons, at
$\nu = \frac{1}{q}$, $q$---odd, \cite{vor3,vor16},
and as a compressible
      fermion sea, at
$q$---even, \cite{vor7,vor11} (the latter is
unstable against BCS-like pairing)
\cite{vor1,vor17}. Assuming
that the CFs are defined by the 1DURs of the
cyclotron subgroup, the hermitian term of this
gauge transformation should be omitted,
because it defines CFs when starting from ordinary
fermions, which are already taken into account in
terms of cyclotron braids.

Let us finally comment on the $\nu=\frac{1}{2}$
state (Hall metal) from the point of view of the
braid approach.
  Within Jain's model,
two flux-tubes attached to composite fermions
completely cancel an external
magnetic field in the mean field approximation (in
other words, the hermitian Ch-S field associated
with Jain's model cancels, in mean field,
  the external magnetic field), and this results
in a Fermi sea, called the Hall metal state
\cite{halperin}. Within Read's approach to
composite particles at $\nu=\frac{1}{2}$,
the complete cancellation of charge takes place
due to the charge density depletion of the vortex
with $q=2$.
  Mutual interchange of 2-fold vortices produces
$q^2\nu \pi=2\pi$ phase shift and including
additional $\pi$ due to electrons, the complexes
of 2-fold vortices with electrons
behave like fermions (without charge)---thus form
a Fermi sea (Hall metal). The instability of the
Fermi system, results next in a paired state
expressed by
the Pfaffian factor, restoring incompressibility
due to the pairing-gap
(BCS-like paired state at $\nu = 5/2$
\cite{vor17,vor9}, also considered for $\nu = 1/2$
and $1/4$ \cite{kr,kr1}).
As Pfaffian \cite{vor1} contributes with $-\pi$ to
the phase shift due to particle interchanges, the
total phase shift of the wave function with
the Jastrow polynomial $\prod_{i>j}(z_i-z_j)^2$
\cite{vor1,vor17} is $\pi$. This phase is given by
the 1DUR of the cyclotron braid group (with $p=3$,
as such a cyclotron braid subgroup
corresponds to the range $\nu \in [1/3,1)$)
assigned by
$p\alpha=3\frac{1}{3}\pi=\pi$, i.e.,
$\alpha=\frac{1}{3}\pi$.
   The representation   ($p=3$,
$\alpha=\frac{1}{3}\pi$) induces the fermion
statistics phase shift of the many-particle wave
function for $\nu =1/2$,
  and in terms of braid-composite fermions,
it corresponds to a net composite electron Fermi
sea (since two loops take away the total external
flux),
   in consistence with the
   local gauge transformation
with $q=2$, thus reproducing fermions (starting
from ordinary fermions) \cite{vor3,vor16}.

\section{Conclusions}

In summary, we argue that, at fractional LL
fillings, braid trajectories must be multi-looped,
while those with lower number
of loops  (including single-looped)  are excluded
due to too short cyclotron radius. This
unavoidable property of braids recovers Laughlin correlations  in a
natural way for
 2D charged systems upon strong magnetic field
 and
explains the structure of CFs both with flux-tubes
or vortices. Classical cyclotron trajectories corresponding
to braids with additional loops (as for fractional
LL fillings) are also multi-looped and this
property explain the
true nature of effective models of 
 flux-tubes and
 vortices. Flux
tubes attached to CFs do not actually
exist  and they only  mimic additional
cyclotron loops in the case of $\nu=\frac{1}{p}$
($p$ odd). Out of the filling fraction
$\nu=\frac{1}{p}$ ($p$ odd), the assumption on
integer number of  flux quanta attached to
particles
in order to create 
 CFs is, however, not  clearly justified
 and only postulated in a heuristic manner.  The introduced 
cyclotron braid approach allows  for avoiding this postulate 
related to CF structure,  including  correction of mapping of FQHE onto IQHE and leading to recovery of 
LL filling hierarchy in a slightly modified version. Nevertheless, in order to rescue Jain's composite fermion structure with rigid integer flux quanta attached to each particle even outside $\nu =\frac{1}{p}$, one has to conjecture the relation between braid cyclotron picture and real cyclic movement of wave-packets represented particles and embracing by their orbits quantized fluxes. 
It leads to the eight-shape multi-looped quasiclassical trajectories in the case when the resulting Jain's field is oriented oppositely to the external field.
The change of direction of cyclotron rotation would be measured 
in the experiment with cyclotron focusing of 2D carriers passing a narrow slot, 
by asymmetry of focusing to the left and to the right with respect to the source slot, when passing $\nu=\frac{1}{2}$ via changing the magnitude of the external field.

  Unitary representations of cyclotron braids
allow also for a self-consistent explanation of
compressible states at fillings with even
denominators.
For example,
$\nu =1/2$ metal Hall state corresponds to
composite anyons with $ p\alpha=3 \frac{1}{3}
\pi=\pi$ signature of 1DUR of the $p=3$ cyclotron
braid subgroup.



\bibliographystyle{model1-num-names}
\bibliography{abbr,braid}








\end{document}